# Machine Learning Assisted Inertia Estimation using Ambient Measurements

Mingjian Tuo, *Student Member, IEEE* and Xingpeng Li, *Senior Member, IEEE*

*Abstract*—With the increasing penetration of converter-based renewable resources, different types of dynamics have been introduced to the power system. Due to the complexity and high order of the modern power system, mathematical model-based inertia estimation method becomes more difficult. This paper proposes two novel machine learning assisted inertia estimation methods based on long-recurrent convolutional neural (LRCN) network and graph convolutional neural (GCN) network respectively. Informative features are extracted from ambient measurements collected through phasor measurement units (PMU). Spatial structure with high dimensional features and graphical information are then incorporated to improve the accuracy of the inertia estimation. Case studies are conducted on the IEEE 24-bus system. The proposed LRCN and GCN based inertia estimation models achieve an accuracy of 97.34% and 98.15% respectively. Furthermore, the proposed zero generation injection bus based optimal PMU placement (ZGIB-OPP) has been proved to be able to maximize the system observability, which subsequently improves the performance of all proposed inertia estimation models.

## I. INTRODUCTION

Renewable energy sources (RES) are replacing traditional synchronous generators with the primary goal of carbon dioxide emission reduction and environmental benefits [1]. Increasing penetration of inverter-based resources such as wind power, solar photovoltaics (PV) and energy storage systems (ESS), has degraded the system inertia during this transition and introduced different dynamics into traditional power systems [2]. Traditionally, power system inertia plays an important role in regulating rate of change of frequency (RoCoF) and frequency excursion after a disturbance. Insufficient system inertia would lead to dramatical change in frequency and further results in under frequency load shedding as well as tripping of generator protection devices; the failure of successive units would furthermore cause cascading outages [3].

Inertia estimation can ensure the accountability and reliability of inertia response through implementation of frequency control ancillary services [4]-[5]. In large-scale deregulated interconnection power systems, inertia information is only available within operators' own territories. Thus, system-wide inertia estimation is important for operators to provide frequency regulation services. Traditionally, system frequency response is analyzed by looking at the collective performance of all generators using a system equivalent model. Based on event measurements and mathematical model, the system inertia could be estimated by the number and size of actively connected synchronous units.

Reference [6] proposed an inertia estimation approach which divides the system into multiple subareas and estimates inertia of each subarea separately, but the approximation made in mathematical model introduces additional errors. The Electric Reliability Council of Texas (ERCOT) uses a real-time sufficiency monitoring tool to monitor inertia based on the operating plans submitted by the generation resources [7]. As previous inertia estimation methods are based on mathematical models, the ability of these methods is dependent on factors such as the size of disturbance, accuracy of frequency measurement and location of measurement point relative to in-feed loss [8]. They are difficult to implement in inertia monitoring due to randomness of system events. Therefore, inertia estimation using ambient wide area measurements are considered as a more accurate method that can reflect the system-wide status.

Inertia estimation method based on mathematical model is highly dependent on accuracy of measurements from phasor measurement units (PMUs) or equivalent devices. However, modern power systems are connected to different devices which provide frequency regulation service, and inertia constant estimation purely based on synchronous generators is inaccurate [9]. Moreover, RES and other inverter-based sources are traditionally considered passive in terms of inertial response. The variability nature of RES also imports uncertainties into the system inertial response as well as system inertia constant [10]. Recent study in [11] shows that control schemes emulating synchronous machine response can be used to contribute system inertia. Therefore, the swing equation-based models may not be able to capture the entire characteristics. In addition, nonlinearities in system frequency response such as deadbands and saturations cannot be taken into considerations either. Thus, the estimated value based on mathematical model may suffer inaccuracy in various conditions.

The invention and development of PMU based wide area measurements systems (WAMS) enable the application of data-driven techniques in power system analysis [12]. A neural network-based inertia estimation technique is proposed in [13], which utilizes inter-area model information as neural network inputs and estimates the inertia constant as an output of the network. However, this approach only estimates the

Mingjian Tuo and Xingpeng Li are with the Department of Electrical and Computer Engineering, University of Houston, Houston, TX, 77204, USA (e-mail: mtuo@uh.edu; Xingpeng.Li@asu.edu).

inertia constant for large systems with only traditional synchronous generation. A convolutional neural network (CNN) based model is proposed in [14], which estimates the system inertia through frequency response and RoCoF data. Graphs are a kind of data structure which models a set of objects (nodes) and their relationships (edges) [15]. Recent advances in deep neural network (DNN) offer an opportunity to integrate graph topology into a neural network, creating a graph neural network (GNN) model [16]. Power system can be represented as a graph with high dimensional features and interdependency among buses. This perspective may offer a better state of the art machine learning for power systems analysis.

To bridge aforementioned gaps, we propose two model-free ambient measurements-based machine learning approaches in this paper to dynamically estimate the system inertia constant. The major contributions of this work are as follows,

1) First, although data driven approaches have been investigated for system inertia estimation in previous work [13], the topological information and high dimensional features haven't been studied thoroughly. To tackle this issue, a long-term recurrent convolutional network (LRCN) based algorithm are proposed to efficiently process temporal measurements, and a graph convolutional neural networks (GCN) assisted method model is used to efficiently identify spatial data. The proposed models have been compared with other state of art methods such as CNN and DNN models.
2) Secondly, previous inertia estimation methods only utilize frequency and RoCoF data derived from single frequency measurement, results may suffer high errors when non-monotonic frequency deviation occurs. We consider measuring heterogeneous responses from multiple nodal PMUs, and the extracted features are then reformulated into the form of graph structure for GCN training.
3) Thirdly, a wrapper feature selection algorithm is used to optimize the feature combination set for inertia estimation. The proposed ambient measurements-based algorithms are examined under multiple noise conditions, which demonstrates the proposed methods can improve the estimation accuracy as well as estimator robustness.
4) Last, a zero generation injection bus based optimal PMU placement (ZGIB-OPP) method is proposed in this paper to maximize the observability of WASM, which furthermore improves the performance of all inertia estimation models.

The remainder of this paper is organized as follows. In section II, the frequency dynamics of power systems are described. Section III details the proposed inertia estimation algorithms using LRCN and GCN techniques. Section IV describes the simulation setup, and the results and analysis are presented in Section V. Section VI presents the concluding remarks and future work.

## II. SYSTEM FREQUENCY DYNAMICS

The frequency of the power system is one of the most important metrics that indicate the system stability. Traditionally, the frequency is treated as unique of the whole power system, which is derived from the system equivalent model extended from one-machine swing equation.

The inertia constant of a generator is a parameter describing the ability of synchronous generator in counteracting the frequency excursion due to power imbalance occurring in power systems. The energy stored in large rotating generator and some industrial motors gives them the tendency to remain rotating. The rotational energy $E_i$ in the rotor of the machine at nominal speed is defined by the following formula:

$$E_i = \frac{1}{2} J_i \omega_i^2 \quad (1)$$

where $J_i$ is the moment of inertia of the shaft in kg·m²s and $\omega_i$ is the nominal rotational speed. The inertia constant $H_i$ is then given in seconds, which can be expressed as:

$$H_i = \frac{J_i \omega_i^2}{2 S_{B_i}} \quad (2)$$

where $S_{B_i}$ is the generator rated power in MVA. When multiple generators connected to the power system, dynamics of these generators' rotors are directly coupled with the grid electrical dynamics. Thereby the power system could be represented by a single equivalent model of inertia. The total power system inertia $E_{sys}$ is then considered as the summation of the kinetic energy stored in all dispatched generators synchronized with the power system. It can be shown in the form of either the stored kinetic energy or inertia constants as follows.

$$E_{sys} = \sum_{i=1}^{N} \frac{1}{2} J_i \omega_i^2 = \sum_{i=1}^{N} H_i S_{B_i} \quad (3)$$

The inertia constant of the power system in seconds is given by the equation below,

$$H_{sys} = \frac{\sum_{i=1}^{N} H_i S_{B_i}}{S_B} \quad (4)$$

where $S_B$ is the total rated power of the whole system.

The simplified system equivalent model is based on the extension of one-machine swing equation. For a single machine, the dynamic of its rotor can be described in (5) with $M = 2H$ denoting the normalized inertia constant and $D$ denoting damping constant respectively.

$$\Delta P_m - \Delta P_e = M \frac{d\Delta\omega}{dt} + D\Delta\omega \quad (5)$$

where $\Delta P_m$ is the total change in mechanical power and $\Delta P_e$ is the total change in electric power of the power system. $d\Delta\omega/dt$ is commonly known as RoCoF.

The dynamics between power and frequency during a short period of time following a disturbance can be modeled by the swing equation. As an approximation, the equation commonly used for system equivalent model is expressed as follows,

$$\frac{d\Delta\omega}{dt} = \frac{\Delta P_m - \Delta P_e}{2 H S_B} \omega_n \quad (6)$$

If the size of disturbance and RoCoF are accurately measured in advance, then the total inertia of the system can be estimated based on the equivalent expression. However, such approximation fails to capture the entire characteristics, non-linearities in system frequency response and controls are not taken into consideration.

The power system with frequency control loops is shown in Fig 1. For primary frequency control, once the power mismatch event has occurred and frequency has started to drop from nominal value, the deviation is fed into closed control droop where turbine-governor counteracts the power mismatch [17]. In this paper, we only focus on the short period following the disturbance. Thus, the model can be simplified further when we analyze only primary frequency control, i.e. model the dynamics before any secondary control gets involved.

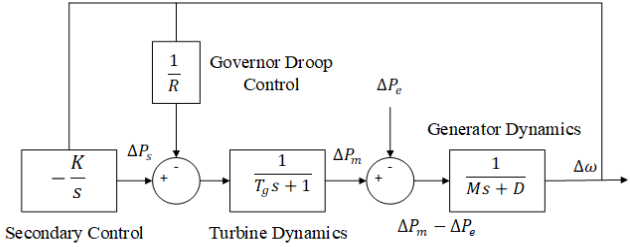

Fig. 1. Generator transfer function model.

As most inertia estimation approaches rely on event transient measurement of collective system model following recorded disturbances, study in [18] have found that the approximation may introduce high error due to inertia heterogeneity, and thus cause issues in system operations.

Different from focusing on the collective performance of the power system equivalent model, the frequency response experienced by each bus could be very distinct. Therefore, dynamic model is preferred in modern power system analysis. Using the topological information and system parameters, when multiple generators connected in a bus, equivalent equation (5) can be extended and applied to all buses to describe the oscillatory behavior of each individual bus,

$$m_i \ddot{\theta}_i + d_i \dot{\theta}_i = p_{in,i} - p_{e,i} \quad (7)$$

where $m_i$ and $d_i$ denote the inertia coefficient and damping ratio for node $i$ respectively, while $p_{in,i}$ and $p_{e,i}$ refer to the power input and electrical power output, respectively. Accordingly, the ambient measurements of nodes from the system would provide more information of each subarea and thus improve the accuracy of estimation model.

## III. INERTIA ESTIMATION

### A. Wide Area Monitoring System

Synchronized measurement technology makes it possible to sample analogue voltage and current wave data in synchronism with a global positioning system (GPS) clock, and record the corresponding frequency related data from widely distributed locations.

PMUs are widely used for modern power systems. Fig.2 shows the topology of wide area monitoring system (WAMS). Measurements from PMUs are obtained from widely distributed locations, and synchronized with respect to a GPS clock. Synchrophasor technologies allow direct measurement of frequency and bus voltages. With the development of PMU based WAMS, the accuracy of measurements improves significantly.

Most PMUs can calculate up to 30 to 60 samples per cycle with the GPS time stamp provided by hardware that has an accuracy of millisecond or higher [12], reporting rates of 10 - 240 samples per second are allowed. In this paper, the sampling rate of PMU is set to 200 per second.

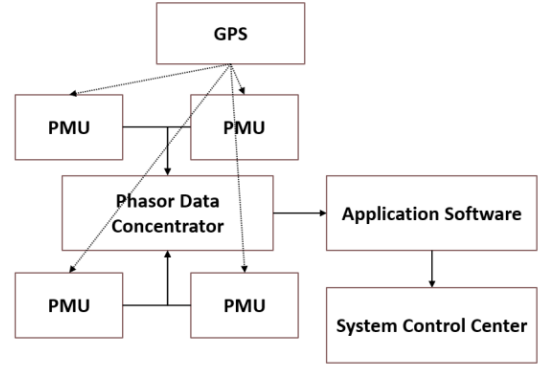

Fig. 2. Wide area measurement system.

### B. System Perturbation using Probing Signal

Low level probing signal method has been conventionally used for generator dynamic studies. A modified form of detrended fluctuation analysis has been introduced to determine the event suitability for probing signal method [14]. With PMUs installed throughout the system providing highly accurate measurements and test improvement such as microperturbation method (MPM), the probing signal method can provide effective approach for system inertia identification without affecting system stability [19].

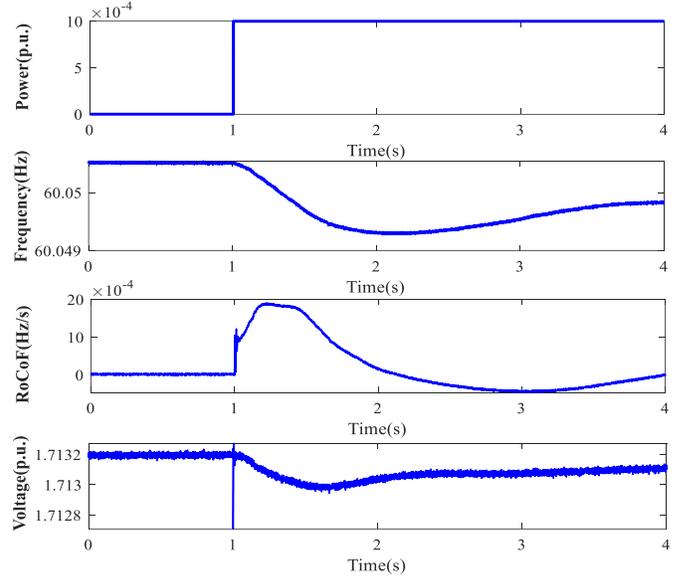

Fig. 3. A sample of probing signal, ambient measurements for $P_E$=0.001 p.u.

A sample probing signal, fed to the system with an amplitude of $P_E$, and corresponding PMU measurements are shown in Fig. 3. With varying system inertia and probing signal amplitude, a number of ambient measurements of $\Delta\omega$, $\Delta\dot{\omega}$ and $v$ can then be collected.

### C. Inertia Estimation using LRCN

Motivation of CNNs roots in the history of neural networks for graph data processing, recurrent neural networks (RNN) are utilized on graphs and cycles. Study in [20] has shown that CNNs have the ability to extract spatial features and compose them to construct expressive representations. An example of a convolutional neural network is shown in Fig. 4 [20]-[21].

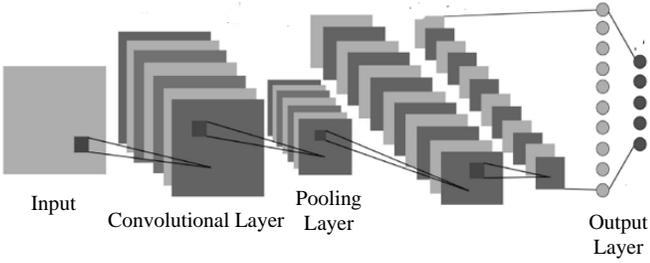

Fig. 4. Illustration of convolutional neural network architecture [20]-[21].

Long short-term memory (LSTM) is an extended frame of RNN which can exhibit temporal behavior of time-series input data. An LSTM cell typically compromises three gates: input, forget and output gates [22].

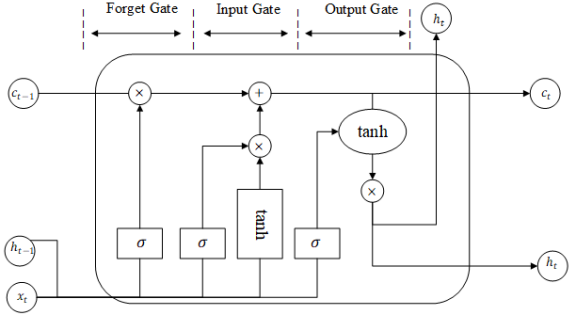

Fig. 5. Illustration of an LSTM cell.

The fundamental equations of LSTM network can be represented as follows:

$$h_t = (1 - z_t) * h_{t-1} + z_t * h_t \quad (8)$$
$$z_t = \sigma(W_f[h_{t-1}, x_t] + b_f) \quad (9)$$

where, $x_t$ is the network input; $h_t$ is the output state of the neuron from LSTM network; $h_{t-1}$ is the previous state of the neuron; $z_t$ computes the necessary information and removes the irrelevant data; $\sigma$ is the sigmoid function; $W_f$ is the weight and $b_f$ is the bias.

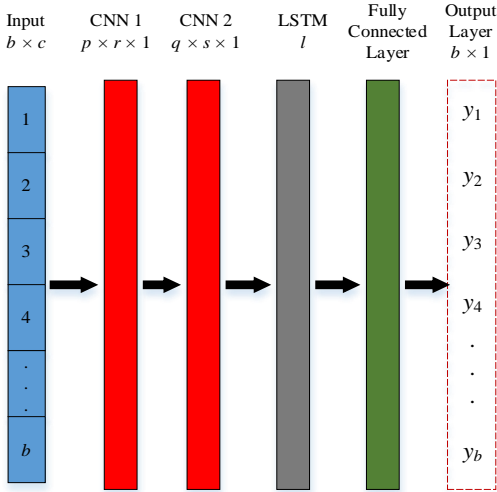

Fig. 6. General architecture of proposed LRCN model.

Since LRCN leverages the strength of rapid progress in CNN and has the ability to capture the dependencies in a sequence, it has been successfully used in computer vision, image processing, and other fields in signals and time-series analysis [23]. The architecture of the proposed LRCN model is illustrated in Fig. 6 and can be trained to estimate system inertia from ambient measurements obtained from the PMU.

The proposed LRCN model first processes the measurements input with 1-D-CNN layers, whose outputs are then fed into the LSTM recurrent sequence model; and the fully connected layer finally produces the estimated inertia constant. The samples in training set are defined in batches which will be propagated through the networks. One epoch of training is completed when all the training samples have been passed forward and backward once. The number of iterations is defined as the number of passes, and each pass uses the same batch size that is the number of samples. At each training iteration, the LRCN model input size is $b \times c$, and the output will be a column vector of size $b$ with inertia estimates for corresponding input in the time period. While the dimension of $c$ is determined by the set of features and feature sampling rate.

The mean squared error (MSE) measures the average squared difference between actual and predicted outputs. The goal of training is to minimize MSE via back propagation which will provides best estimator [24]. The fully connected network used in this model includes one flatten layer and two hidden layers. MSE is defined as:

$$MSE = \frac{1}{n}\sum_{i=1}^{n}(y_i - \tilde{y}_i)^2 \quad (10)$$

where $n$ is the total number of training samples, $y_i$ is the actual value of $i^{th}$ output, and $\tilde{y}_i$ is the estimated value corresponding to the $i^{th}$ output. Similarly, the weight update equation via back propagation is expressed as:

$$w_{t+1} = w_t - \alpha \frac{\partial E_{MSE}}{\partial w_t} \quad (11)$$

where $w_t$ is the weight for current iteration, $w_{t+1}$ is the updated weight for next iteration, $\alpha$ is the learning rate, and $E_{MSE}$ is the MSE obtained from expression (11).

### D. Graph Convolutional Neural Networks

Geometric deep learning is a recent emerging field. Traditional CNNs have limitations in processing graphical data which have explicit topological graph correlation embedded [25]. Recent advancement of CNN results in the rediscovery of GNNs. GCN has been developed by extending the convolution operation onto graphs and in general onto non-Euclidean spaces. Previous studies in [26]-[27] have proved that GCN provides state-of-arts performance in graph analysis tasks.

Power system is an interconnected network of generators and loads. The graph structure of the power system consists of nodes (buses) and edges (branches). The branches in the power system are undirected, such graphs provide information on buses and their connections. The convolution operator in propagation module is used to aggregate information from neighbors. Considering $\mathcal{G} = (\mathcal{V}, \mathcal{E})$ as an undirected graph representing a power system, where $\mathcal{V} \in \mathbb{R}^N$ denotes its nodes and $\mathcal{E} \in \mathbb{R}^K$ denotes its edges. Let $A \in \mathbb{R}^{N \times N}$ be the adjacency matrix of $\mathcal{G}$, we can define a renormalization trick as:

$$V = \tilde{D}^{-\frac{1}{2}} \tilde{A} \tilde{D}^{-\frac{1}{2}} \quad (12)$$

where $\tilde{A} = A + I_N$ represents an adjacency matrix with self-connections. Typically, the element at $(i, j)$ of the adjacency matric $A$ is defined as follows:

$$A_{ij} = \begin{cases} 1; & \text{if } \mathcal{V}_i, \mathcal{V}_j \in \mathcal{V}, (\mathcal{V}_i, \mathcal{V}_j) \in \mathcal{E} \\ 0; & \text{if } \mathcal{V}_i, \mathcal{V}_j \in \mathcal{V}, (\mathcal{V}_i, \mathcal{V}_j) \notin \mathcal{E} \end{cases} \quad (13)$$

where $(\mathcal{V}_i, \mathcal{V}_j)$ denotes the branches from $i$ to $j$. The diagonal degree matrix $\tilde{D}$ for $\mathcal{G}$ is defined as $\tilde{D}_{ii} = \sum_j \tilde{A}_{ij}$.

The graph convolutional layer is defined as follows,

$$F^l(X, A) = \sigma\left(V F^{(l-1)}(X, A) W_k^l + b^l\right) \quad (14)$$

where $F^l$ is the convolutional activations and $b^l$ is the bias matrix at the $l$-th layer; $F^0 = X$ is the input matrix. Fig. 7 demonstrates the message passing mechanism in forward propagation, a target node (bus 8) receiving information from its neighboring nodes.

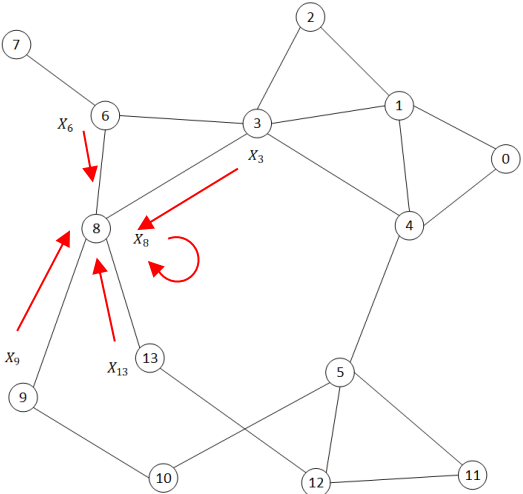

Fig. 7. Example of message passing mechanism.

### E. Optimal PMU Placement

The cost of PMUs depends on the count of measurement channels available [28]. Due to the high cost of having a PMU at each node, various techniques are used to solve the optimal PMU placement (OPP) problem for observability of the system for static and dynamic state estimators [29].

Traditionally OPP is a binary optimization problem, the objective functions considered in OPP problems in previous studies are mainly for minimization of the number of PMUs (15), maximization of observability (16), or both as a multi-objective function.

$$Min\ F_1 = \sum_{i=1}^{N} x_i \quad (15)$$

$$Max\ F_2 = \sum_{i=1}^{N} o_i \quad (16)$$

where $x_i, i = 1, 2, \ldots, N$ is elements of the PMU installation indication row vector $X$. For the base case, $o_i$ denotes the observability of each bus.

The complete topological observability can be expressed as follows,

$$O = A \cdot X \quad (17)$$
$$O \geq u \quad (18)$$

where $u$ is a row vector $N \times 1$ consisting of binary variables, representing that the monitored bus is observed by PMU.

## IV. SIMULATION SETUP

### A. Overview

The IEEE 24-bus system [30] was used for the experiment to collect the training data. The system has 24 buses (17 buses with loads), 38 branches, and 38 generators. The system inertia $M$ typically ranges from 3s to 8s. Hence, to ensure the practicality of the proposed model, the measurements snapshots were collected for 11 different values of $M$ from 3s to 8s with an increment of 0.5s following daily dispatches considering the RES penetration. Similarly, probing signals with 100 different values of $P_E$ from 0.001 p.u. to 0.01 p.u. with an increment 0.0001 p.u. were used.

The modeling and simulation of the power system, along with data collection, were conducted in MATLAB/Simulink 2019b. The data pre-processing was conducted in both MATLAB and Python. The proposed LRCN and GCN based models were developed in Python using Keras and PYG.

### B. Data Preprocessing

The initial data analyzed in this study were acquired from PMUs with a sampling rate larger than 200 Hz; nodal measurements of $\Delta\omega$ and $\Delta\dot{\omega}$ are obtained. By using only one second sampling frame for normalization, the real-time applicability of this method is maintained. Similarly, following the same pattern we obtained nodal voltage measurement $v$. Since the training data come from the ambient measurements of all PMUs may suffer different sampling rate which is larger than 200 Hz, without dimension reduction process the original training data would increase the complexity of the model and may also lead to overfitting. In addition, there are bad data in the raw measurements which would introduce high error in analysis results. Therefore, the bad data points are first discarded from raw measurements, and then we downsample the measurements to 200 Hz for next step [14]. Additionally, Gaussian noise signal is added to the constituent tonic to mimic the noisy measurements. Different signal-to-noise ratios (SNR) are investigated in this paper. The data are collected between multiple sessions, all the measurements are normalized by employing min/max normalization between [0, 1].

### C. Feature Selection

To find the best time frame of data extraction, different time windows of the ambient measurements are determined: (1) the time frame is first chosen from 0s to 1s following the perturbation, where initial RoCoF is included; (2) the second time frame is from 0.5s to 1.5s after the signal infeed. With $\Delta\omega$ and $\Delta\dot{\omega}$ as basic features combination, the coefficient of determination and validation accuracy are used as evaluating metrics.

In order to determine the optimal combination of features for inertia estimation model training, a wrapper feature selection method is utilized: (i) the proposed LRCN model is used as the inertia estimator, (ii) accuracy score is used as the evaluation metric, (iii) greedy forward selection as the subset selection policy. The specific metrics for feature evaluations is expressed as follows.

$$ACC = \frac{|\{y \in TD|\ |y - \tilde{y}| \leq \mu\}|}{|TD|} \quad (19)$$

where $ACC$ denotes the proportion of the correctly predicted values with $\mu$ tolerance. $TD$ denotes the validation dataset, and $\mu$ is the predetermined threshold.

### D. PMU Formulations

Constraints of PMU locations are not considered in the base case. With a resampling rate of 200Hz, measurements on each PMU node gives 200 data points at a sampling frame of 1s. In this paper, ambient measurements of $\Delta\omega$, $\Delta\dot{\omega}$ and $v$ on generator buses are assumed as available measurements.

Availability of PMU data are affected by several realistic factors, which have been studied in [8]. Given limited resources, the objective function (16) of OPP is applied in this work to maximize the system observations.

Zero injection bus effect is also added. Traditionally, zero injection bus (ZIB) means no load or generator is connected to it. Since we are more interested in the statement of generators, the impact of zero generation injection bus (ZGIB) is considered in this work where generator is the only factor. The topological observability constraint of each of the bus connected to a ZGIB is updated by introducing virtual connections to every other non-ZGIB bus connected to that same ZGIB.

$$\tilde{o}_i = \sum_{j}^{N} A_{ij}x_j + \sum_{j}^{N} w_{ij}x_j \ \forall i = 1,2,\ldots,N \quad (20)$$

where $w_{ij}$ is an auxiliary binary variable: 1, if buses $i$ and $j$ are both connected to the same ZIB; otherwise 0.

*E. Hyperparameters Selection*

For the convolution layers in the LRCN model, the channel number are set $p = 10$ and $q = 20$, and kernels with sizes $r = s = 3$. Rectified linear unit (ReLU) is used as the activation function. The memory unit value $l$ of LSTM layer is set as 32. The training was operated in batches of 32 data points.

The GCN model consists of one GNN layer and two hidden fully connected layer. The dimension of input features is defined based on the number of available PMUs which gives $n \times d$, where $d$ is the nodal features of GCN layer. The fully connected layers are set as 64 and 128 respectively.

An MSE based dynamic learning rate strategy is used for the training. Learning rate schedule is applied in the training process by reducing the learning rate accordingly, and the factor by which the learning rate will be reduced is set 0.5 and the patience value is set 50 epochs.

## V. RESULT ANALYSIS

A total of 1,100 samples were collected, the entire dataset is first divided into two subsets: 880 samples (80%) for training and 220 samples (20%) for validation.

*A. Time Frame Selection*

To investigate the best time period for feature extraction, all PMUs are considered available in base case. Measurement $\Delta\dot{\omega}$ is selected as training feature combination; training data extracted from two periods are then fed into the aforementioned models, including the proposed LRCN and GCN models.

Table I
Performance Comparison for Different Models

| Models | Period (0.0-1.0s) | | Period (0.5-1.5s) | |
|---|---|---|---|---|
| | Validation Accuracy | MSE | Validation Accuracy | MSE |
| DNN | 93.27% | 0.065 | 78.18 % | 0.314 |
| CNN | 94.46% | 0.052 | 80.36% | 0.236 |
| LRCN | 96.89% | 0.028 | 82.57% | 0.215 |
| GCN | 97.48% | 0.022 | 83.19% | 0.203 |

Performances of different inertia estimation models are summarized in Table I. As it can be observed, the DNN model has the lowest validation accuracy for both scenarios. CNN based model has a relatively higher validation accuracy, showing advantage in processing spatial data. The proposed LRCN model has a validation accuracy of 96.89% with a tolerance of 0.5s for features extracted from period 0.0 - 1.0s period, while it is only 82.57% for the use of features extracted from period 0.5 - 1.5s. GCN based model has the highest validation accuracy in both scenarios at 98.15% and 83.19% respectively. On the whole, the features extracted from the time frame following the disturbance contain prominent inertial response information, and accordingly have a positive impact on the overall performance of inertia constant estimation model.

Fig. 8 compares the distribution of absolute prediction error for the LRCN model with features extracted from 0.5s - 1.5s and 0.0s - 1.0s respectively. Using features extracted from period 0.0 - 1.0s, the coefficient of determination of the proposed LRCN model is 0.9625, which is higher the use of features extracted from period 0.5 - 1.5s at 0.7619.

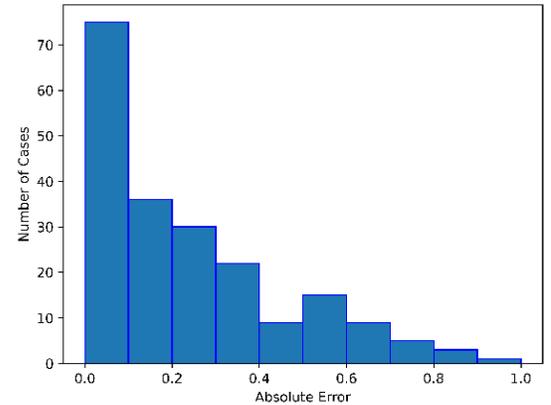

(a) Features extracted from 0.5s - 1.5s.

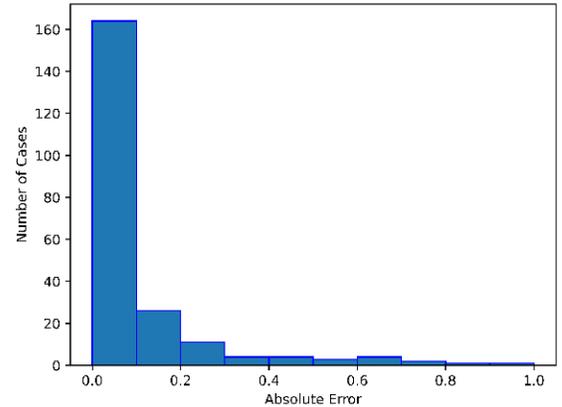

(b) Features extracted from 0.0s - 1.0s.

Fig. 8. Absolute error of prediction with the proposed LRCN model using features extracted from different time periods.

*B. Analysis of Model Performance*

Optimal combination of features extracted from 0.5s - 1.5s period is first selected through greedy forward selection. Table II compares the results with LRCN model as performance estimator. It can be observed that only considering $\Delta\omega$ measurement as input feature provides a validation accuracy of 80.30%. Combination of $\Delta\omega$ and $\Delta\dot{\omega}$ has a highest validation accuracy at 97.34% with 0.5s tolerance, which outperforms other feature combinations. Thus, $\Delta\omega + \Delta\dot{\omega}$ is selected as the optimal combination for model training.

Table II
Comparison of Different Features Sets for LRCN Model

| Features Set | $\Delta\omega$ | $\Delta\dot\omega$ | $\Delta\omega + \Delta\dot\omega$ | $\Delta\omega + \Delta\dot\omega + v$ |
|---|---|---|---|---|
| Validation Accuracy | 80.30% | 96.89% | 97.34% | 95.76% |
| MSE | 0.296 | 0.032 | 0.025 | 0.030 |
| Coefficient of Determination | 0.8945 | 0.9585 | 0.9725 | 0.9564 |

Fig. 9 and Fig. 10 depict the evolution of MSE losses on the training and validation datasets over the training process of the proposed models. As it can be seen, for both LRCN and GCN cases, MSE decreases as the number of epochs increases. In terms of MSE, the validation loss of LRCN model drops faster and reaches minimal value at 0.025. It should be noted that a sudden drop is observed in GCN model training, and the GCN based model has a lower validation loss at 0.020.

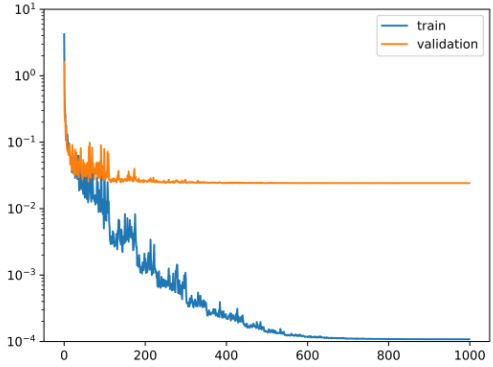
Fig. 9. The learning curve of the proposed LRCN model: MSE losses versus the number of epochs.

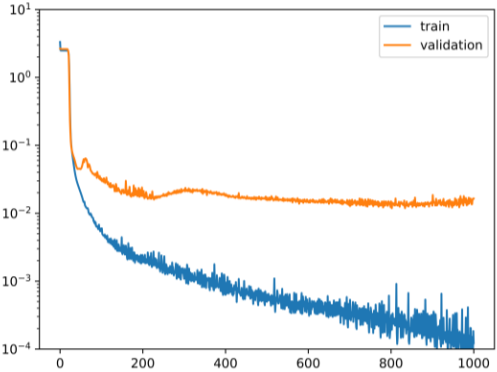
Fig. 10. The learning curve of the proposed GCN model: MSE losses versus. the number of epochs.

The proposed LRCN and GCN based approaches are then compared with benchmark algorithm [14] in Table III, $\Delta\omega$ and $\Delta\dot\omega$ are used as primary features to train the model. Both algorithms are employed to train the inertia constant estimation model. The results show that the CNN model has a validation accuracy of 95.18% with 0.5s tolerance, which is higher than the validation accuracy of DNN model at 93.45%. These results reflect that CNN based model has the better capability to process spatial data comparing to traditional DNN model. Nevertheless, the proposed LRCN model has a validation accuracy of 97.34% with 0.5s tolerance, and GCN model has a validation accuracy at 98.15%. Additionally, the coefficient of determinations of LRCN model and GCN model are 0.9725 and 0.9826 respectively, which are higher than the benchmark CNN and DNN models. An explanation could be that the proposed LRCN and GCN models are more efficient algorithms to identify critical temporal information and graphical information embedded in the collected power system data. Combining Table III and Figs. 9-10, we can observe that GCN has the highest validation accuracy and the lower MSE, indicating that GCN has a better performance in processing graphical data.

Table III
Comparison of Different Models with Optimal Feature Combination

| Model | Validation Accuracy | Coefficient of Determination | MSE |
|---|---|---|---|
| DNN | 93.45% | 0.9224 | 0.058 |
| CNN | 95.18% | 0.9369 | 0.045 |
| LRCN | 97.34% | 0.9725 | 0.025 |
| GCN | 98.15% | 0.9826 | 0.020 |

### C. Impact of Low SNR

In addition to the ideal condition, the proposed models are compared with the benchmark models under high noise conditions. Study in [31] has shown that a SNR of 45dB is considered as a good approximation of noise power under realistic conditions.

Table IV
Comparison of Different models with SNR at 45dB

| Model | w/o SNR | w/ SNR at 45dB |
|---|---|---|
| DNN | 93.45% | 90.84% |
| CNN | 95.18% | 92.13% |
| LRCN | 97.34% | 93.25% |
| GCN | 98.15% | 93.87% |

Table IV shows the inertia estimation accuracy of all models with combination of $\Delta\omega$ and $\Delta\dot\omega$ as training features. After adding additional Gaussian noise signal with a SNR of 45dB to the ambient measurements, the overall MSE increases for both models, while the validation accuracy reduces accordingly. Understandably, a significant reduction in validation accuracy can be observed in both cases.

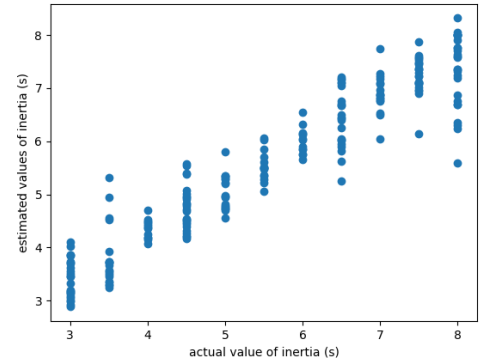
Fig. 11. Prediction results of the benchmark CNN model with SNR at 45dB.

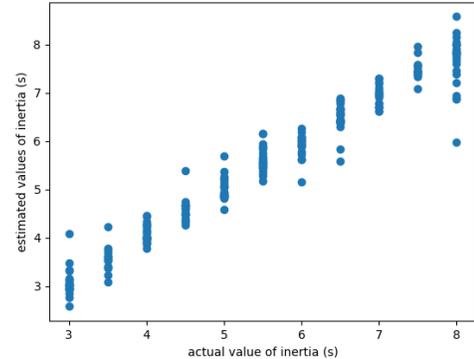
Fig. 12. Prediction results of the proposed LRCN model with SNR at 45dB.

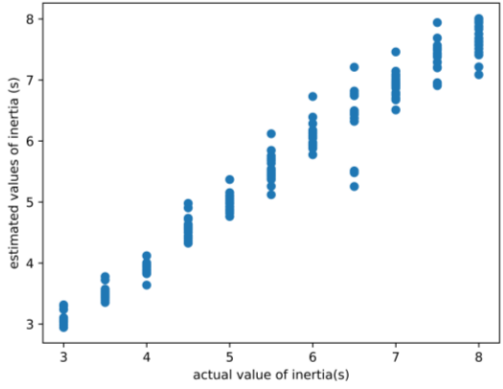

Fig. 13. Prediction results of the proposed GCN model with SNR at 45dB.

The scatter points of CNN model, the proposed LRCN and GCN models when SNR of 45dB is applied are depicted in Figs. 11-13. Results show that only considering measurements of frequency related data may suffer high error when noise is applied.

The method described in this research uses a wrapper feature selection process and then. Results of different feature combinations are listed in Tables V and VI. It can be observed that the proposed LRCN model has a validation accuracy of 93.25% with 0.5s tolerance under the condition of SNR at 45dB. With voltage measurements added, the validation accuracy is improved to 93.87% with 0.5s tolerance.

Table V
Comparison of Different Features Sets for LRCN Model when Considering SNR at 45 dB

| Features Set | $\Delta\omega + \Delta\dot{\omega}$ | $\Delta\omega + \Delta\dot{\omega} + v$ |
|---|---|---|
| Validation Accuracy | 93.25% | 93.76% |
| MSE | 0.119 | 0.098 |
| Coefficient of Determination | 0.9032 | 0.9156 |

Table VI
Comparison of Different Features Sets for GCN Model when Considering SNR at 45 dB

| Features Set | $\Delta\omega + \Delta\dot{\omega}$ | $\Delta\omega + \Delta\dot{\omega} + v$ |
|---|---|---|
| Validation Accuracy | 93.87% | 94.56% |
| MSE | 0.088 | 0.071 |
| Coefficient of Determination | 0.9227 | 0.9449 |

In summary, (i) under a low noise condition, measurements of $\Delta\omega$ and $\Delta\dot{\omega}$ are the optimal set of features suitable for inertia estimation; (ii) with SNR of 45dB added, the performance of benchmark model decreases significantly, while the proposed LRCN model based on optimal features combination of $\Delta\omega$, $\Delta\dot{\omega}$ and $v$ shows higher robustness and better performance.

*D. Optimal PMU Placement*

The application of proposed ZGIB based OPP approach to IEEE 24-bus system is carried, and the corresponding impact on inertia estimation models are investigated in this section.

Table VII listed the results of PMU location considering the proposed ZGIB-OPP method. Given the total number of PMUs limited by two, the proposed OPP method considering ZGIB suggests locating PMUs at buses 2 and 16 for the best inertia estimation performance. When the total number of PMUs increases, the proposed ZGIB OPP adds more buses into the optimal set, indicating the consistency in optimal PMUs locations.

Table VII
Optimal Locations of PMUs given Limited Resources

| No. of PMUs | 2 | 3 | 4 | 5 |
|---|---|---|---|---|
| Bus # | 2, 16 | 2, 16, 21 | 2, 16, 21, 23 | 2, 13, 16, 21, 23 |

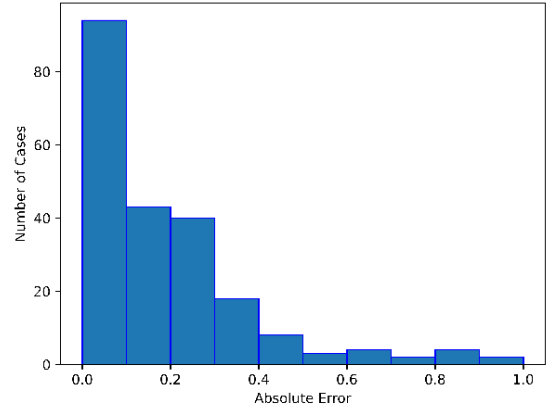

Fig. 14. Distribution of absolute values for random PMU locations.

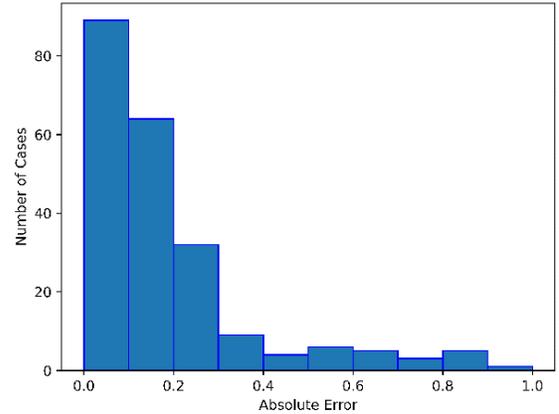

Fig. 15. Distribution of absolute values for ZGIB-OPP.

Table VIII
Comparison of Different Models with Five PMUs Limit

| PMUs Locations | ZGIB-OPP | Random PMUs Location | Full PMUs |
|---|---|---|---|
| DNN | 91.36% | 90.00% | 93.45% |
| CNN | 92.73% | 90.45% | 95.18% |
| LRCN | 95.26 | 92.36% | 97.34% |
| GCN | 95.89% | 94.65% | 98.15% |

We apply the proposed ZGIB-OPP to IEEE 24-bus system, and the system wide measurements of limited channels are then obtained for model training. Fig.14 and Fig.15 compare the distribution of prediction absolute error for different PMUs settings using LRCN model. Comparing to the random PMU location case, the proposed ZGIB-OPP improves the performance of LRCN model, resulting in more samples with absolute error landed within 0.1-0.2s.

Results listed in Table VIII also show that with limited PMU channels, as well as combination of $\Delta\omega$ and $\Delta\dot{\omega}$ as input feature, the validation accuracy of all models drops accordingly. For DNN and CNN model, the validation accuracy drops to 90.00% and 90.45% respectively. Validation accuracy of LRCN and GCN model drops slightly when there are limited channels or under random PMU locations. With ZGIB-OPP applied, GCN based model has the highest accuracy of 95.89%. It should be noted that applying ZGIB-OPP improves the performances of all models, indicating that the overall observability of the WAMS is improved through the proposed ZGIB-OPP algorithm.

## VI. Conclusions

Neural networks have been applied for inertia estimation as extensive amounts of data can be obtained through power system digital equipment and advanced measuring infrastructures such as PMUs. In this paper, LRCN and GCN based learning algorithms are proposed to estimate system inertia constant. System wide ambient measurements based on WAMS are used as candidate features for model training, and a wrapper feature selection is also used to optimize the feature combination. Considering the limitation on PMU settings, an ZGIB-OPP method is proposed to maximize the observability of the WAMS given limited PMU resources. Results indicate that the proposed LRCN and GCN models have better performances than the benchmark DNN and CNN models. The proposed LRCN model and GCN model also show high robustness under conditions with higher noises. The proposed ZGIB-OPP method has been proved to be capable of improving the performance of all implemented models. Considering that the IEEE 24-bus system model used in this research has a mix generation of both synchronous generators and inverter-based resources, the proposed approach can also be applied to estimate inertia constant in realistic conditions.

## References

[1] Mingjian Tuo and Xingpeng Li, "Long-Term Recurrent Convolutional Network-based Inertia Estimation using Ambient Measurements," 2022 IEEE Industry Applications Society Annual Meeting (IAS), 2022, pp. 1-6.

[2] Mingjian Tuo, and Xingpeng Li, "Dynamic Estimation of Power System Inertia Distribution Using Synchrophasor Measurements", in *Proc. 52$^{nd}$ North Amer. Power Symp.*, April. 2021, virtually, Tempe, AZ, USA, pp. 1–6.

[3] F. Milano, F. Dörfler et al., "Foundations and challenges of low-inertia systems," in *2018 Power Systems Computation Conference (PSCC)*, Jun. 2018, pp. 1–25.

[4] S. You, Y. Liu, J. Tan, M. T. Gonzalez, X. Zhang, Y. Zhang and Y. Liu, "Comparative Assessment of Tactics to Improve Primary Frequency Response Without Curtailing Solar Output in High Photovoltaic Interconnection Grids," IEEE Trans. Sustainable Energy, vol.10, Issue 2, pp. 718-728, 2019.

[5] Gu, H., Yan, R., & Saha, T. K. "Minimum synchronous inertia requirement of renewable power systems." *IEEE Transactions on Power Systems.* 33(2), 1533–1543, June. 2017.

[6] Y. Cui, S. You, and Y. Liu, "Ambient Synchrophasor Measurement Based System Inertia Estimation," in *2020 IEEE Power Engineering Society General Meeting (PESGM),* 2020: IEEE, pp. 1-5.

[7] ERCOT, "Inertia: Basic Concepts and Impacts on the ERCOT Grid," ERCOT, Tech. Rep., Apr. 2018.

[8] Phillip M. Ashton, Christopher S. Saunders, Gareth A. Taylor, Alex M. Carter and Martin E. Bradley, "Inertia Estimation of the GB Power System Using Synchrophasor," *IEEE Transactions on Power Systems*. vol. 30, no. 2, pp. 701-709, March. 2015.

[9] Federico Milano, and Álvaro Ortega, "A Method for Evaluating Frequency Regulation in an Electrical Grid – Part I: Theory," *IEEE Trans. Power Syst.*, vol. 36, no. 1, pp. 183-193, Jan. 2021.

[10] Svenska Kraftn¨at, Statnett, Fingrid and Energinet.dk, "Challenges and Opportunities for the Nordic Power System," Tech. Rep., 2016.

[11] Bala Kameshwar Poolla, Saverio Bolognani and Florian Dörfler, "Optimal Placement of Virtual Inertia in Power Grids," *IEEE Transactions on Automatic Control*. vol. 62, no. 12, pp. 6209-6220, Dec. 2017.

[12] R.B. Sharma and G.M. Dhole. Wide area measurement technology in power systems. Procedia Technology, 25:718–725, 2016.

[13] A. Schmitt and B. Lee, "Steady-state inertia estimation using a neural network approach with modal information," *2017 IEEE Power Energy Society General Meeting*, Jul. 2017, pp. 1–5.

[14] Abodh Poudyal, Robert Fourney, Reinaldo Tonkoski, Timothy M. Hansen, Ujjwol Tamrakar, and Rodrigo D. Trevizan, "Convolutional Neural Network-based Inertia Estimation using Local Frequency Measurements", in *Proc. 52$^{nd}$ North Amer. Power Symp.*, April. 2021, virtually, Tempe, AZ, USA.

[15] F. Scarselli, M. Gori, A. C. Tsoi, M. Hagenbuchner and G. Monfardini, "The Graph Neural Network Model," in *IEEE Transactions on Neural Networks*, vol. 20, no. 1, pp. 61-80, Jan. 2009.

[16] Thuan Pham and Xingpeng Li, "Reduced Optimal Power Flow Using Graph Neural Network", 54th North American Power Symposium, Salt Lake City, UT, USA, Oct. 2022.

[17] Markovic, U., Chu, Z., Aristidou, P., et al.: 'LQR-based adaptive virtual synchronous machine for power systems with high inverter penetration', IEEE Trans. Sust. Energy, 2019, 10, (3), pp. 1501–1512

[18] A Adrees, JV Milanović and P Mancarella, "Effect of Inertia Heterogeneity on Frequency Dynamics of Low-inertia Power Systems," *IET Generc., Transmiss. Distrib.*, vol. 13, no. 14, pp. 2951–2958, Jul. 2019.

[19] J. Zhang and H. Xu, "Online identification of power system equivalent inertia constant," *IEEE Transactions on Industrial Electronics*, vol. 64, no. 10, pp. 8098–8107, Apr. 2017.

[20] Sridharan V, Tuo M, Li X, "Wholesale electricity price forecasting using integrated long-term recurrent convolutional network model." Energies 15(20):7606

[21] Saleh Albeiwi and Ausif Mahmood, "A Framework for Designing the Architectures of Deep Convolutional Neural Networks" *Entropy*, 19(6), 242, May 2017.

[22] S. Hochreiter and J. Schmidhuber, ''Long short-term memory,'' Neural Comput., vol. 9, no. 8, pp. 1735–1780, 1997.

[23] Jeff Donahue, Lisa Anne Hendricks, Marcus Rohrbach, Subhashini Venugopalan, Sergio Guadarrama, Kate Saenko, and Trevor Darrell, "Long-term Recurrent Convolutional Networks for Visual Recognition and Description," " in IEEE Transactions on Pattern Analysis and Machine Intelligence, vol. 39, no. 4, pp. 677-691, April 1, 2017.

[24] Akshita Chugh, "MAE, MSE, RMSE, Coefficient of Determination, Adjusted R Squared - Which Metric is Better?". https://medium.com/analytics-vidhya/mae-mse-rmse-coefficient-of-determination-adjusted-r-squared-which-metric-is-better-cd0326a5697e.

[25] Pope, P.E.; Kolouri, S.; Rostami, M.; Martin, C.E.; Hoffmann, H. Explainability Methods for Graph Convolutional Neural Networks. In Proceedings of the IEEE Conference on Computer Vision and Pattern Recognition (CVPR), Long Beach, CA, USA, 16–20 June 2019.

[26] F. Ebrahimzadeh, M. Adeen, and F. Milano, "On the impact of topology on power system transient and frequency stability," in Proc. IEEE Int. Conf. Environ. Electr. Eng., IEEE Ind. Commercial Power Syst. Eur. (EEEIC/I&CPS Europe), Jun. 2019, pp. 1-5.

[27] T.Weckesser, H. Jóhannsson, M. Glavic, and J. Østergaard, "An improved on-line contingency screening for power system transient stability assessment," Electr. Power Compon. Syst., vol. 45, no. 8, pp. 852-863, May 2017.

[28] Johnson T and Moger T, "A critical review of methods for optimal placement of phasor measurement units," Int. Trans. on Elec. Ener. Sys. 31.

[29] N. H. Abbasy and H. M. Ismail, "A unified approach for the optimal PMU location for power system state estimation," IEEE Trans. Power Syst., vol. 24, no. 2, pp. 806–813, May 2009.

[30] Mingjian Tuo, Arun Venkatesh Ramesh, Xingpeng Li, "Benefits and Cyber-Vulnerability of Demand Response System in Real-Time Grid Operations", *IEEE Smart Grid Comm*, Nov. 2020, Tempe, AZ, USA.

[31] M. Brown, M. Biswal et al., "Characterizing and quantifying noise in PMU data," in *2016 IEEE Power and Energy Society General Meeting (PESGM)*, Jul. 2016, pp. 1–5.